\documentstyle[12pt]{article}

\newcommand{\gsim}{\stackrel{\mbox{\raisebox{-0.1ex}{\scriptsize $>$}}}
{\mbox{\raisebox{-0.5ex}{\scriptsize $\sim$}}}}

\begin{document}
\newlength{\myleftmargin}
\newlength{\paperwidth}
\setlength{\paperwidth}{210mm}
\setlength{\myleftmargin}{20mm}
\setlength{\oddsidemargin}{30pt}
\setlength{\evensidemargin}{\myleftmargin}
\setlength{\textwidth}{400pt}
\setlength{\textheight}{600pt}
\setlength{\headheight}{5pt}
\setlength{\topmargin}{5pt}
\pagestyle{plain}

\begin{flushright}
KOBE-FHD-95-04 \vspace{2ex} \\
April 1995 \vspace{2ex} \\
\end{flushright}

\begin{center}
\renewcommand{\thefootnote}{\fnsymbol{footnote}}
 {\large \bf Neutrino Oscillations by Interaction with Moduli}\\
\vspace{4em}
K. Kobayakawa${}^{1}$
\footnote
{On leave from Faculty of Human Development, Kobe University},
Y. Sato${}^{2}$ and S. Tanaka${}^{3}$ \\
\vspace{3em}
${}^{1}$
        {\it Fukui University of Technology, }\\
        {\it Gakuen-cho, Fukui 910, Japan}\\

\vspace{2em}
${}^{2}${\it Graduate School of Science and Technology, }\\
        {\it Kobe University, Nada, Kobe 657, Japan}\\
\vspace{2em}
${}^{3}$
        {\it Division of Natural Environment and High Energy Physics, }\\
        {\it Faculty of Human Development, }\\
        {\it Kobe University, Nada,Kobe 657, Japan}\\
\end{center}
\vspace{3em}

\begin{center}
\begin{abstract}
\baselineskip=20pt
We would like to point out the possibility to detect
the low-energy signals of moduli in the superstring theory
in the neutrino oscillation.
The idea is based on the characteristics that the couplings of moduli
are different in matter. We estimate
the oscillation probability both in the long baseline
and in the solar neutrino oscillations
and examine the detectable region of the moduli effect.
\end{abstract}
\end{center}

\clearpage
\baselineskip=16pt

\vspace{1em}

Recent LEP data \cite{LEP}
suggest the evidence of Grand Unified Theories
such as SU(5), SO(10), flipped SU(5) and so on. Furthermore
the data fit better on including supersymmetry.
On the theoretical side to solve the gauge hierarchy problem
the idea of supersymmetry is very persuasive.
However, SUSY GUT does not contain the interaction of gravity.
At present it is conceived that the superstring theory alone
may include all interactions consistently in the theory.
Phenomenologically the heterotic superstring theory
\cite{Gross} is most attractive.
There are several ways of compactification and
after that there come out very many vacua \cite{Green}.
They are parametrized, in general, by moduli \cite{Ferrara}
which are singlet superfields under the
gauge group of the standard model,
$\rm{SU(3)_{C}}\times \rm{SU(2)_{L}}\times \rm{U(1)_{Y}}$. For example,
some of them describe the size and shape of the compactified space.
But masses of moduli are not known, though their vacuum expectation values
are supposed to be of the
order of Planck scale. Their interactions with matter are also
model-dependent. Even the number of moduli depend on the structure
of the vacuum considered.
Consequently it is very helpful to detect the moduli.

In this paper we would like to point out that moduli characteristic
in the superstring theory may give new low-energy signals which could
be tested in the neutrino oscillation experiments.
Moduli generally couple to ordinary matter with nonrenormalizable
interactions. Such couplings are expressed in
the superpotential effectively as (in the lowest dimension)
\begin{equation}
P_{nonren} = \frac{c^{I}_{ijk}}{M_{\rm{S}}}\varphi_{i}\varphi_{j}
\varphi_{k}M_{I},~~~~~~~ (I=1,2,3,\dots),
\end{equation}
where $\varphi_{i,j,k}$ are matter superfields,
$M_{I}$ are moduli superfields and $M_{\rm{S}}$ is the string scale
($\sim 10^{18}\rm{GeV})$. Such terms at low energies induce
Yukawa-type couplings between the ordinary matter
and (real) scalar fields or pseudoscalar fields i.e.
moduli:
\begin{eqnarray}
{\cal L}_{Y} &=& \frac{<H_{2}>}{M_{\rm{S}}}h_{ij}^{(\nu)}
\bar{\nu}_{R}^{i}\nu_{L}^{j}M_{I} +
\frac{<H_{2}>}{M_{\rm{S}}}h_{ij}^{(u)}
\bar{u}_{R}^{i}u_{L}^{j}M_{I} \nonumber \\
&+& \frac{<H_{1}>}{M_{\rm{S}}}h_{ij}^{(d)}
\bar{d}_{R}^{i}d_{L}^{j}M_{I} +
\frac{<H_{1}>}{M_{\rm{S}}}h_{ij}^{(\ell)}
\bar{\ell}_{R}^{i}\ell_{L}^{j}M_{I} + h.c.,
\label{Yukawa}
\end{eqnarray}
where $i$ and $j$ are generation indices($i=1,2,3$) and $<H_{1,2}>$
are the vacuum expectation values of the Higgs doublets
and $\gamma$-matrices are dropped.
While dilaton $S$ interacts with ordinary matter universally
like graviton, moduli interact (or not interact) with various couplings.
Moduli interact with ordinary matter as a coherent attractive force
\cite{Cvetic}.
(We also consider the possibility of moduli interaction
generating a repulsive force.)
Since the interaction strength is comparable to that of
gravity force, this
behaves as a kind of fifth force if the mass of the exchanged particle
is small enough \cite{Cvetic}\cite{Macorra}.
Ordinary moduli mass is expected to be of the order of the gravitino mass.
Because moduli have a flat potential perturbatively and they must get mass
by nonperturbative effects. It is usually said that it occurs after
supersymmetry breaking.
So it would be as heavy as other scalar sparticles.
However, there are arguments that some moduli would
have very tiny mass:

(1) for real $M_{I}$ the moduli mass ($m_{M_{I}}$) may be
induced by radiative corrections
($m_{M_{I}}\simeq 10^{-18} \rm{GeV}$) \cite{Cvetic},
or there may be a special cancellation of two terms in the mass equation
\cite{Carlos}. In ref.\cite{Kelley}, it is estimated that $m_{M_{I}}$
can be about $m_{\frac{3}{2}}^{2}/ReM_{I}$, where
$m_{\frac{3}{2}}$ is the gravitino mass.

(2) for imaginary $M_{I}$, in ref.\cite{Macorra} it is argued
that $m_{M_{I}}$ can be $2\times10^{-24}\rm{GeV}$.
However, in ref.\cite{Kelley} it is said that they are massless.
In ref.\cite{Ibanez}, on the other hand, they are said to gain huge
mass of the order of the SUSY-breaking scale.

There is no definite mass
which can be calculated numerically, and also the form of the scalar
potential is not known yet. So
we do not get into the details of the models here and take a mass of a
modulus (especially tiny one) as a parameter $m_{M_{I}}$ and its
relative interaction
strength as parameters $f_{ij}$, and explore the possibilities
of finding the effects of moduli in the terrestrial experiments,
not in cosmology.

\vspace{1em}

First we discuss long baseline neutrino oscillations.
The planning experiments such as from FNAL to
SOUDAN2 \cite{FNAL} is illustrated
schematically in Fig.1.
The neutrino source from an accelerator is
located at the point $x_{1}$, and the detector at the point $x_{2}$.
The muon neutrino ($\nu_{\mu}$) beam with energy $E$
(of the order from one GeV to a few ten GeV) propagates along the $x$-axis.
We assume, for simplicity, that there is at least one modulus
which interacts with $\nu_{\tau}$
and/or $\nu_{\mu}$ and u- or d-quark (or electron).
For example, $h_{22}^{(\nu)}\neq 0, h_{11}^{(u)}\neq 0$ and
others can be zero in eq.(\ref{Yukawa}).
Although the interaction strength is gravitational, it may be detectable
in the neutrino oscillations
when $m_{M_{I}}$ is very tiny.
We take it in this paper in the range of
$10^{-24}-10^{-15}$GeV.

We define the eigenstate of mass plus moduli interaction as
$(\nu_{2}, \nu_{3})$, and
the flavor eigenstate as $(\nu_{\mu}, \nu_{\tau})$.
The latter eigenstate is expressed by the former with a mixing angle
$\zeta$ as
\begin{equation}
\left(
\begin{array}{l}
\nu_{\mu} \\
\nu_{\tau}
\end{array}
\right)
=
\left(
\begin{array}{cc}
\cos\zeta      & \sin\zeta\\
-\sin\zeta     & \cos\zeta
\end{array}
\right)
\left(
\begin{array}{l}
\nu_{2} \\
\nu_{3}
\end{array}
\right).
\label{eqn:mixing}
\end{equation}

The Hamiltonian of mass eigenstate is given by
\begin{equation}
H =
\left(
\begin{array}{cc}
p+\frac{m_{2}^{2}}{2p}-f_{22}\phi      & -f_{23}\phi\\
-f_{32}\phi                           & p+\frac{m_{3}^{2}}{2p}-f_{33}\phi
\end{array}
\right),
\label{eqn:Hamiltonian}
\end{equation}
where $p$ is the momentum of a neutrino beam and
$m_{2}$ and $m_{3}$ are the masses
of mass eigenstates.
$f_{ij}\phi$ in eq.(\ref{eqn:Hamiltonian}) represent the potentials
induced by moduli interaction.
The difference of modulus coupling with matter is stuffed
into $f_{ij}(|f_{ij}|\leq 1; i,j=2,3)$.
We can take $f_{23}=f_{32}$.
Since $f_{23}$ is not necessarily zero, this Hamiltonian is not
always diagonal.
In a relativistic case $\phi$ is represented as the product
\cite{Gasperini}
of energy of a neutrino beam and
the potential per unit mass due to moduli interaction with matter
\begin{eqnarray}
\phi &=& EV, \nonumber \\
V &=& G_{M} \frac{M}{r}\exp(-\mu r).
\label{eqn:exponential}
\end{eqnarray}
Here $\mu$ can be regarded as almost same as the moduli mass
and $G_{M}$ is a common coupling constant of moduli so that
the maximum value among $|f_{ij}|$ is unity.
In eq.(\ref{eqn:exponential}),
$M$ is the mass of the matter which is interacting
with neutrino interchanging moduli.

To estimate $V$, the contribution to $V$ from the whole earth is added up.
Then
\begin{eqnarray}
\phi_{global} &=& -\frac{3G_{M}M_{E}E}{2\mu^{2}R_{E}^{3}}
\frac{R_{E}+\mu^{-1}}{\xi_{0}} \nonumber \\
& \times & [
\exp\{-\mu(R_{E}-\xi_{0})\} - \exp\{-\mu(R_{E}+\xi_{0})\}]
+ \frac{3G_{M}M_{E}E}{\mu^{2}R_{E}^{3}},
\label{eqn:global}
\end{eqnarray}
where $M_{E}$ and $R_{E}$ denote the mass and the radius of the earth,
respectively. $\xi_{0}$ is the distance
between the beam and the center of the earth
as shown in Fig.1. We can put
$\xi_{0} = R_{E}$ approximately.
The value of eq.(\ref{eqn:global}) is almost the same as that
of the sphere with radius $\mu^{-1}$ because of the exponential
decrement in eq.(\ref{eqn:exponential}).

The Hamiltonian (\ref{eqn:Hamiltonian}) can be diagonalized and we get
\begin{equation}
i\frac{d}{dx}
\left(
\begin{array}{l}
\nu_{2} \\
\nu_{3}
\end{array}
\right)
=
\left(
\begin{array}{cc}
p+\frac{m_{2}^{2}}{2p}+ \alpha    & 0\\
0     & p+\frac{m_{3}^{2}}{2p}+ \beta
\end{array}
\right)
\left(
\begin{array}{l}
\nu_{2} \\
\nu_{3}
\end{array}
\label{eqn:motion}
\right),
\end{equation}
where $\alpha$ and $\beta$ are parameters introduced for convenience sake:
\begin{eqnarray}
\label{eqn:alpha}
\alpha &=& \lambda_{2}- p - \frac{m_{2}^{2}}{2p},  \\
\label{eqn:beta}
\beta &=& \lambda_{3}- p - \frac{m_{3}^{2}}{2p},   \\
\lambda_{2,3} &=& p+\frac{m_{2}^{2}+m_{3}^{2}}{4p}
-\frac{f_{22}+f_{33}}{2}\phi \nonumber \\
 &\pm& \left[\left(\frac{\Delta m^{2}}{4p}\right)^{2}
+\left(\frac{\Delta f\phi}{2}\right)^{2}
-\frac{\Delta m^{2}}{4p}\Delta f \phi
+f_{23}^{2}\phi^{2}\right]^{2},
\label{eqn:parameter}
\end{eqnarray}
where $\Delta m^{2} = m_{3}^{2}-m_{2}^{2}$
and $\Delta f = f_{33}-f_{22}(|\Delta f|\leq 1)$
represents the difference of the coupling constants of
the two neutrino species with matter.

Solving eq.(\ref{eqn:motion}) and using eqs.(\ref{eqn:mixing})
and (\ref{eqn:alpha})-(\ref{eqn:parameter}),
we obtain the oscillation probability:
\begin{eqnarray}
\lefteqn{
P(\nu_{\mu}\rightarrow\nu_{\tau})}\nonumber \\
&=& \sin^{2}2\zeta \nonumber \\
& \times & \sin^{2}\left[
\frac{L}{c}\left\{
\left(\frac{\Delta m^{2}}{4E}
\right)^{2} +
\left(\frac{\Delta f\phi}{2}
\right)^{2}
- \frac{\Delta m^{2}}{4E}\Delta f \phi
+ f_{23}^{2}\phi^{2}
\right\}^{\frac{1}{2}}
\right].
\label{eqn:oscillation}
\end{eqnarray}
The first term inside the brace is due to the vacuum
oscillation and the last three terms are due to moduli interaction.
So comparing the two kinds of contributions, we can examine the
effect of moduli interaction.
The former is proportional to $E^{-1}$ and the latter is proportional
to $E$. Therefore the higher the energy of neutrino is, the larger the
effect of moduli is.
The effect of moduli interaction
may be detected experimentally,
if it is at least about $10^{-3}$ of that of vacuum oscillation
\cite{Minakata}.
Here we neglect the term of $f_{23}^{2}\phi^{2}$ so it might be
underestimation of moduli effect.

We examine the detectable region of two parameters
$\mu$ and $\Delta f$ as follows.
The force induced by moduli interactions behaves like
the fifth force which many experiments have tested
and put restrictions.
First fixing the value of $\mu$,
which we set at a reciprocal of the force range $\lambda$, we take
$G_{M}$ in eq.(\ref{eqn:exponential}) at
the maximum value of allowable $G_{5}$.
In this way we get the limit value of $\Delta f$ at each $\mu$.
Denoting $\alpha=\frac{G_{M}}{G_{N}}$, where $G_{N}$ is the
gravitational constant, we use
the limit values of
$(\mu [{\rm GeV}], \alpha)$
for the attractive force from ref.
\cite{Nature}: for example, $(2.0\times10^{-22},3.0\times10^{-6})$,
$(2.0\times10^{-20}, 1.6\times10^{-4})$,
$(2.0\times10^{-18}, 5.0\times10^{-4})$, and so on.
Similarly, for the repulsive force the restrictions are found in
ref.\cite{Stacey}.

Let us consider two versions of $\Delta m^{2}$. First, if $\nu_{\tau}$
is regarded as a candidate of dark matter, then $\Delta m^{2}$ is
expected to be about $100{\rm eV}^{2}$ \cite{Harari}.
Second, according to Kamiokande atmospheric neutrino data
\cite{Kamiokande},
$\Delta m^{2}\simeq 10^{-2}{\rm eV}^{2}$. Next, we take as the energy
of neutrino $E$ the following three typical examples: \\
(i) KEK $\rightarrow$ Kamioka ($E=1.4$GeV) \cite{KEK} \\
(ii) FNAL $\rightarrow$ SOUDAN2 ($E=10$GeV) \cite{FNAL} \\
(iii) $E=1$TeV (such neutrinos are detectable in e.g.DUMAND
\cite{DUMAND})\\
Figs.2a and 2b correspond to
$\Delta m^{2}=100{\rm eV}^{2}$ and $\Delta m^{2}=10^{-2}{\rm eV}^{2}$
in the case that moduli interaction is
an attractive force.
The observable region is the upper part of the dotted line in (i),
the dashed line in (ii)
and the solid line in (iii), respectively.
Namely the lines show the limit $\Delta f
\phi=10^{-3}\frac{\Delta m^{2}}{2E}$.
As the energy of the neutrino increases,
the detectable region becomes wide.
The effect of moduli interaction is more significant
in the case of Fig.2b than that of Fig.2a.
Similarly in the case of repulsive force,
the observable region of $\Delta f$ is shown
in Fig.3a ($\Delta m^{2}=100{\rm eV}^{2}$)
and in Fig.3b ($\Delta m^{2}=10^{-2}{\rm eV}^{2}$).

We will comment on eq.(\ref{eqn:oscillation}) a little more.
The formula in the brace can be written as
\begin{equation}
\frac{L}{c}\left(\frac{\Delta m^{2}}{4E}
-\frac{\Delta f}{2}\phi\right),
\label{eqn:brace}
\end{equation}
for $f_{12}=0$.
In eq.(\ref{eqn:brace}) the first term is
\begin{equation}
\frac{L}{c}\frac{\Delta m^{2}}{4E}=1.27
\frac{\left(\frac{\Delta m^{2}}{{\rm eV}^{2}}\right)}
{\left(\frac{E}{{\rm GeV}}\right)}
\left(\frac{L}{{\rm km}}\right),
\label{eqn:vacuum}
\end{equation}
and the second term is approximately given from
eq.(\ref{eqn:global}) as
\begin{eqnarray}
\frac{\Delta f}{2}\phi &=& \frac{3}{4}\frac{L}{c}\Delta f
\frac{G_{M}ML}{\mu^{3}R_{E}^{3}} \nonumber \\
 &=& 2.54\Delta f\left(\frac{\alpha}{10^{-4}}\right)
\frac{1}{\left(\frac{\mu}{10^{-20}{\rm GeV}}\right)^{2}}
\left(\frac{E}{{\rm GeV}}\right)
\left(\frac{L}{{\rm km}}\right).
\label{eqn:moduli}
\end{eqnarray}
When $\Delta f=1$ and all other physical quantities
are O(1) in the denoted units,
both values of eqs.(\ref{eqn:vacuum}) and
(\ref{eqn:moduli}) are near $\frac{\pi}{2}$
which gives the maximum value of $P(\nu_{\mu}\rightarrow\nu_{\tau})$.
\vspace{1em}

We touch upon the effect of moduli interaction to solar neutrino
oscillations briefly.
Taking MSW effect \cite{MSW}\cite{MSW2} into account
and using the number density
of electrons $N_{e}$ \cite{Bahcall} which reads
\begin{equation}
N_{e}(R)=245N_{A}\exp\left(-10.54\frac{R}{R_{sun}}\right)
\end{equation}

We get the probability that $\nu_{e}$ changes into $\nu_{\mu}$ as
\begin{eqnarray}
\label{eqn:prob}
P(\nu_{e}\rightarrow\nu_{\mu}) &=&
\sin^{2}\zeta_{m}\sin^{2}\left[
\frac{1}{2c}\left\{\frac{\Delta m'^{2}}{2E}L_{sun}
-\Delta f\Phi \right\} \right], \\
\Phi &=& \frac{93\pi G_{M}}{\mu^{2}}Em_{p}
\left(\alpha_{p} + \alpha_{n}
+\frac{m_{e}}{m_{p}}\right)N_{A}R_{sun} \nonumber \\
& & \times \left[\exp\left(-10.54\frac{R_{min}}{R_{sun}}
\right)-\exp(-10.54)\right],
\end{eqnarray}
where $\Delta m'^{2} = m_{2}'^{2}-m_{1}'^{2}$,
$\Delta f=f_{22}-f_{11}$, $\alpha_{p}$ and $\alpha_{n}$
correspond to contribution of protons and neutrons, respectively.
$\nu_{e}$'s are supposed to be generated around the distance of
$R_{min}(\simeq 0.1R_{sun})$ from the
center and $L_{sun}=R_{sun}-R_{min}\simeq 0.9R_{sun}$.
$\Phi$ is the
integrated value of $\phi$ with respect to
$R$ from $R_{min}$ to $R_{sun}$.
The second term inside the brace of eq.(\ref{eqn:prob}) is
the effect of moduli interaction.
The ratio of $|\Delta f \Phi|$ to $\Delta m'^{2}L_{sun}/2E$ is
estimated to be $ 8 \times  10^{-9}$ for a typical example of
$\mu=10^{-22}$GeV,
$G_{M}=10^{-6}G_{N}$, $E=10$eV and $\Delta m'^{2}=10^{-5}{\rm eV}^{2}$.
Thus we conclude that the moduli interaction affords no significant
effect in the solar neutrino oscillation.

\vspace{2em}

There are discussions on direct phenomenological consequences of moduli
in cosmology in ref.\cite{Carlos} and others.
Here we point out that moduli
may give signals even in accelerator experiments, that is, long
baseline neutrino oscillations.
In some cases, the effect might be seen in a short baseline
neutrino oscillation. For example, in the CHORUS experiment
\cite{CHORUS} they try to detect $\tau$ after
$\nu_{\tau}$-nucleon charge current interactions.
At $E\simeq30{\rm GeV}$, $L=0.8{\rm km}$,
$\Delta f \phi/2=61\Delta f$ for $\mu\sim10^{-20}-10^{-24}$GeV.
Then the experiment proves affirmative for
moduli, if $|\Delta f|\gsim10^{-2}$.

In conclusion if neutrino oscillation experiments
will be scrupulously performed
with various conditions, a clue of the form of moduli interaction
with matter and mass of a modulus might be obtained.
When moduli effect is detected in neutrino oscillation,
there must exist at least one modulus with tiny mass and
the structure of the true vacuum or the mechanism of SUSY
breaking would be restricted.

\clearpage

Figure captions
\baselineskip=16pt

Fig.1: Schematic of Long Baseline Neutrino Oscillations. \\
The neutrino beam is injected from the accelerator
at the point $x_{1}$ and detected at the point $x_{2}$.

\vspace{1em}

Figs.2a, 2b: Observable region in the ($\mu$, $\Delta f$) plane
for the case of attractive force. \\
a: $\Delta m^{2}=100{\rm eV}^{2}$;
b: $\Delta m^{2}=10^{-2}{\rm eV}^{2}$. \\
The observable regions are shown by the
upper part of lines. The dotted, the dashed
and the solid lines
correspond to (i)$E=1.4$GeV, (ii)10GeV, (iii)1TeV, respectively.

\vspace{1em}

Figs.3a, 3b: Observable region in the ($\mu$, $\Delta f$) plane
for the case of repulsive force. \\
The observable regions and other notations are the same as Fig.2.
\end{document}